\documentclass[preprint,12pt]{aastex}

\begin{document}
\title{TYPE I PLANET MIGRATION IN NEARLY LAMINAR DISKS}
\shorttitle{Migration in Nearly Laminar Disks}
\author{H.~Li\altaffilmark{1}, S.~H. Lubow\altaffilmark{2}, 
S.~Li\altaffilmark{1}, D.N.C. Lin\altaffilmark{3}}
\altaffiltext{1}{Los Alamos National Laboratory, Los Alamos, NM 87545;
  {\tt hli@lanl.gov; sli@lanl.gov}} 
\altaffiltext{2}{Space Telescope Science
  Institute, 3700 San Martin Drive, Baltimore, MD 21218; {\tt
    lubow@stsci.edu}}
\altaffiltext{3}{UCO/Lick Observatory, University of California, Santa
  Cruz, CA 95064; {\tt
    lin@ucolick.org}}


\begin{abstract}

  We describe 2D hydrodynamic simulations of the migration of low-mass
  planets ($\leq 30 M_{\oplus}$) in nearly laminar disks (viscosity
  parameter $\alpha < 10^{-3}$) over timescales of several thousand
  orbit periods.  We consider disk masses of $1$, $2$, and $5$ times
  the minimum mass solar nebula, disk thickness parameters of $H/r =
  0.035$ and $0.05$, and a variety of $\alpha$ values and planet
  masses.  Disk self-gravity is fully included.  Previous analytic
  work has suggested that Type~I planet migration can be halted in
  disks of sufficiently low turbulent viscosity, for $\alpha \sim
  10^{-4}$.  The halting is due to a feedback effect of breaking
  density waves that results in a slight mass redistribution and
  consequently an increased outward torque contribution.  The
  simulations confirm the existence of a critical mass ($M_{cr} \sim
  10 M_{\oplus}$) beyond which migration halts in nearly laminar
  disks.  For $\alpha \ga 10^{-3}$, density feedback effects are
  washed out and Type~I migration persists.  The critical masses are
  in good agreement with the analytic model of Rafikov (2002).  In
  addition, for $\alpha \la 10^{-4}$ steep density gradients produce a
  vortex instability, resulting in a small time-varying eccentricity
  in the planet's orbit and a slight outward migration.  Migration in
  nearly laminar disks may be sufficiently slow to reconcile the
  timescales of migration theory with those of giant planet formation
  in the core accretion model.

\end{abstract}


\section{Introduction}
\label{sec:intro}

The standard theory of Type~I (low planet mass) migration presents a
challenge for understanding planet formation. According to the core
accretion model, the growth time from planetesimals to gas giant
planets is dominated by a phase that occurs when a newly formed solid
core with mass $\sim 10 M_{\oplus}$ accretes gas (Pollack et al 1996,
Hubickyj et al 2007).  The duration of this slow phase, $\sim 10^6 y$,
is determined by a thermal bottleneck that prevents the gaseous
envelope from contracting, until it achieves sufficient mass. On the
other hand, the standard theory of Type~I migration (Tanaka et al
2002) predicts that planets in the slow growth phase would migrate
into the disk center in $\sim10^5 y$ for the minimum solar mass
nebula.  These migration timescales have been confirmed by
multidimensional hydrodynamical simulations (Bate et al 2003, D'Angelo
\& Lubow 2008).  Recently, Ida \& Lin (2008) and Schlaufman, Lin, \&
Ida (2008) have studied the effect of the ice line on the disk
surface density profile and consequently on the Type I migration. They
obtained much better agreement with the observed extrasolar planet
mass-semimajor axis distribution if the Type I migration is reduced by
an order-of-magnitude from the linear theory values.

The shortness of the standard migration timescale has motivated
investigations of possible effects to slow or even reverse migration.
These include magnetic fields (Terquem 2003), magneto-rotational
instability (MRI) turbulent fluctuations (Nelson \& Papaloizou 2004),
and density traps (Menou \& Goodman 2004).  Recently, protoplanet
migration in the non-isothermal disks has been investigated
(Paardekooper \& Mellema 2006; Baruteau \& Masset 2008; Paardekooper
\& Papaloizou 2008; Kley \& Crida 2008). These simulations show
indications of slowing migration due to coorbital torques for certain
ranges of gas diffusivity and turbulent viscosity.  In this Letter we
discuss another mechanism that naturally occurs when the disk
turbulent viscosity is sufficiently small.

The Tanaka et al (2002) Type~I migration rates were derived under the
assumption that the disk density distribution is unaffected by the
presence of the planet.  Numerical simulations commonly adopt
turbulent viscosity parameter values $\alpha \ga 10^{-3}$. Such values
are suggested by considering the observationally inferred disk masses
and accretion rates for T Tauri stars (e.g., Hartmann et al 1998).
With such $\alpha$ values, turbulent diffusion suppresses disk
disturbances for planets of mass less than $0.1 M_J$.  The numerical
simulations then satisfy the Tanaka et al (2002) assumptions and yield
migration rates that are in close agreement.

The various models of planet formation by core accretion typically
involve lower $\alpha$ values, $\alpha < 10^{-3}$ (see Cuzzi \&
Weidenschilling 2006), which we refer to as nearly laminar values.  To
form planetesimals via gravitational instability from small dust
particles (Safronov 1969, Goldreich \& Ward 1973) requires the dust
layer to be very thin $\sim 10^{-4} H$, suggesting $\alpha \ll
10^{-4}$. For the solids to be dynamically decoupled from the gas
requires a dust layer disk of thickness $\la 10^{-2} H$, again
suggesting nearly laminar conditions. Cuzzi \& Weidenschilling (2006)
estimate that $\alpha \la 2 \times 10^{-4}$ in order that meter size
solids avoid destructive effects of collisions due to turbulent
motions.

In the planet formation regions of disks, considerations of the MRI
(Balbus \& Hawley 1991) suggest that the disk maybe unstable only in
surface layers, due to the low levels of ionization below these layers
(Gammie 1996).  A major uncertainty is the abundance of small grains
that can suppress the instability.  The disk may be nearly laminar for
the purposes of planet formation.  However, surface layer turbulent
fluctuations may propagate disturbances to the disk midplane. They may
provide some effective turbulence in that region as well (Fleming \&
Stone 2003; Turner \& Sano 2008).

In nearly laminar disks, density waves launched by a planet at various
Lindblad resonances can redistribute disk mass as they damp. The disk
turbulent viscosity needs to be sufficiently small for the density
perturbation to not diffuse away.  The redistributed gas slightly
enhances (reduces) the disk density interior (exterior) to the orbit
of an inwardly migrating planet.  Some analytic studies have shown
that such density feedback effects could slow and even halt the migration
(Hourigan \& Ward 1984; Ward \& Hourigan 1989; Ward 1997; Rafikov
2002).  The critical planet mass at which the feedback becomes
important depends on the efficiency of the damping of waves excited by
the planet.  Wave damping may be due to the shock dissipation and will
generally occur nonlocally, at some distance from the radius where the
wave is launched.  In this {\em Letter}, we explore the consequences
of nearly laminar disks on planet migration by means of nonlinear 2D
numerical simulations.

\section{Numerical Method and Initial Setup}
\label{sec:init}

We assume that the protoplanetary disk is thin and can be described by
the 2D isothermal Navier-Stokes equations in a cylindrical \{$r,
\phi$\} plane centered on the star with vertically integrated
quantities.  The differential equations are the same as given in Kley
(1999).  Simulations are carried out using a hydro code developed at
Los Alamos (Li et al.  2005). We also use the local comoving angular
sweep as proposed in the FARGO scheme of Masset (2000) and modified in
Li et al. (2001).  The equations of motion of the planets are the same
as given in D'Angelo et al. (2005), which we adapted to the polar
coordinates with a fourth-order Runge-Kutta solver.  During each
hydrodynamics time step, the motion of the planet is divided into
several substeps so that the planet always moves within 0.05 local
grid spacing, $\delta = [(\Delta r)^2 + (r_i\Delta \phi)^2]^{1/2}$, in
one substep.  The disk gravitational force on the planet is assumed to
evolve linearly with time between two hydrodynamics time steps.
Furthermore, we have implemented a full 2D self-gravity solver on our
uniform disk grid (Li, Buoni, \& Li 2008).  This solver uses a mode
cut-off strategy and combines FFT in the azimuthal direction and
direct summation in the radial direction.  The algorithm is
sufficiently fast that the self-gravity solver costs less than $10\%$
of the total computation cost in each run. This code has been
extensively tested on a number of problems. With our pseudo-3D
treatment (see Li et al. 2005 for details) and a small (a few grid
size) softening distance in the planet's potential, migration rates from
simulations with sufficient viscosity (dimensionless kinematic
viscosity $\nu \simeq 10^{-6}$) agree well (within a few percent) with
the 3D linear theory results by Tanaka et al. (2002). As the softening
distance increases to $r_H$, the migration rates from such simulations
are $\sim 30\%$ slower than the 3D linear theory result.  The runs
presented here use $r_H$ as the softening distance.

The 2-D disk is modeled between $0.4 \leq r \leq 2$. The planet is
initially located at $r=1$, which corresponds to a physical distance
of Jupiter's orbital radius (5.2 AU), and orbits about a $1 M_{\odot}$
star. The unit of time is the initial orbit period $P$ of the planet,
which is about 12 yr.  A corotating frame that rotates with the
initial angular velocity of the planet is used. The coordinate plane
is centered on the central star at $(r,\phi)=(0,0)$ (acceleration due
to frame rotation is also included, the so-called indirect term).  The
disk is assumed to be isothermal throughout the simulated region,
having a constant sound speed $c_s$.  The dimensionless disk thickness
is scaled by the initial orbital radius of the planet
$h=c_s/v_{\phi}(r=1)$, where $v_{\phi}$ is the Keplerian velocity.  We
consider values $h = 0.035$ or $0.05$ in the simulations. We have also
made runs using constant disk aspect ratio $H/r$, and found that our
main conclusions are not changed.   

Our numerical schemes require two ghost cells in the radial direction
(the angular direction is periodic).  Holding these ghost cells at the
initial steady state values produced the weakest boundary reflections
among all boundary conditions we investigated.  We choose an initial
surface density profile normalized to the minimum mass solar nebular
model (Hayashi 1981) as $\Sigma(r) = 152\, f (r/5{\rm AU})^{-3/2}$ gm
cm$^{-2}$, where $f$ ranges from $1-5$ in our simulations.  The
initial rotational profile of the disk is calculated so that the disk
will be in equilibrium with the disk self-gravity and pressure
(without the planet).  The mass ratio between the planet and the
central star is $\mu=M_{p}/M_{\ast}$, which ranges from $3\times
10^{-6}$ to $10^{-4}$.  The planet's Hill (Roche) radius is $r_H = r_p
(\mu/3)^{1/3}$.  The dimensionless kinematic viscosity $\nu$ (normalized by
$\Omega^2 r$ at the planet's initial orbital radius) is taken to
be spatially constant and ranges between $0$ and $10^{-5}$. For
$h=0.05$, the effective Shakura and Sunyaev $\alpha = \nu /h^2$ at the
initial planet radius ranges between $0$ and $4\times10^{-3}$. We have
performed various tests to show that when $\nu=0$, the effective
numerical viscosity in our simulations is $\nu < 10^{-9}$ or $\alpha <
4\times 10^{-7}$.  We typically evolve the disk without the planet for
10 $P$. Subsequently, the planet's gravitational potential is gradually
``turned-on'' over a 30-orbit period, allowing the disk to respond to
the planet potential gradually. Note that the time shown in all the
figures in this {\it Letter} starts at the time of the planet release.
Runs are made typically using a radial and azimuthal grid of
$(n_{r}\times n_{\phi})= 800\times 3200$, though we have used higher
resolution to ensure convergence on some runs. Simulations typically
last several thousand orbit periods at $r=1$.

\section{Results}

\begin{center}
\begin{figure}
\centerline{\includegraphics[height=3in,width=5in,angle=0]{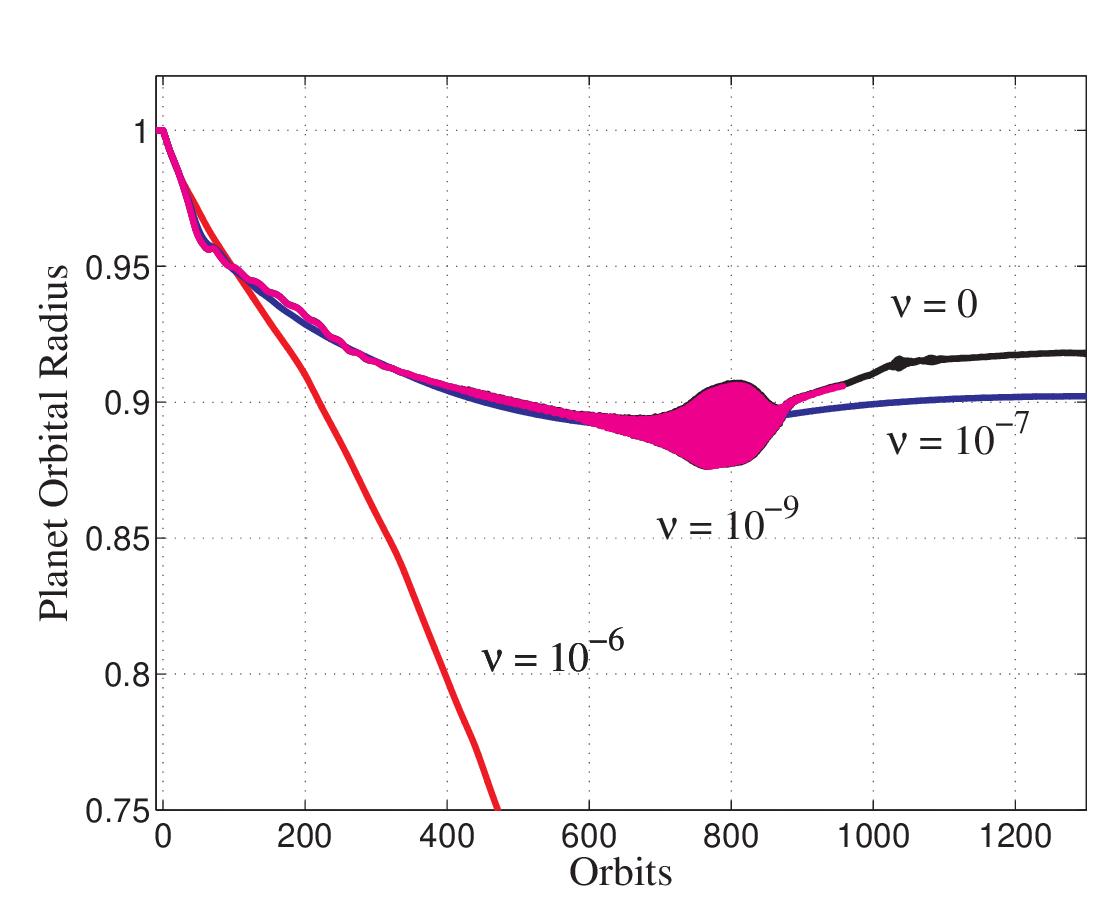}}
\caption{Influence of dimensionless disk viscosity $\nu$  on the
  migration  of a planet with
  mass $10 M_{\oplus}$ in a disk with $h = 0.035$ and $f=5$. 
  The vertical axis is the orbital radius in units of the initial orbital
  radius. The horizontal axis is the time in units of the initial planet
  orbital period, $P$.
  For dimensionless kinematic viscosity 
  $\nu = 10^{-6} $ ($\alpha = 8 \times 10^{-4}$), the planet undergoes
  the typical Type I migration,
  but its migration is slowed  significantly when $\nu$ is smaller.  
  \label{fig:vis_all}}
\end{figure}
\end{center}

\begin{center}
\begin{figure}
\centerline{\includegraphics[height=3in,width=5in,angle=0]{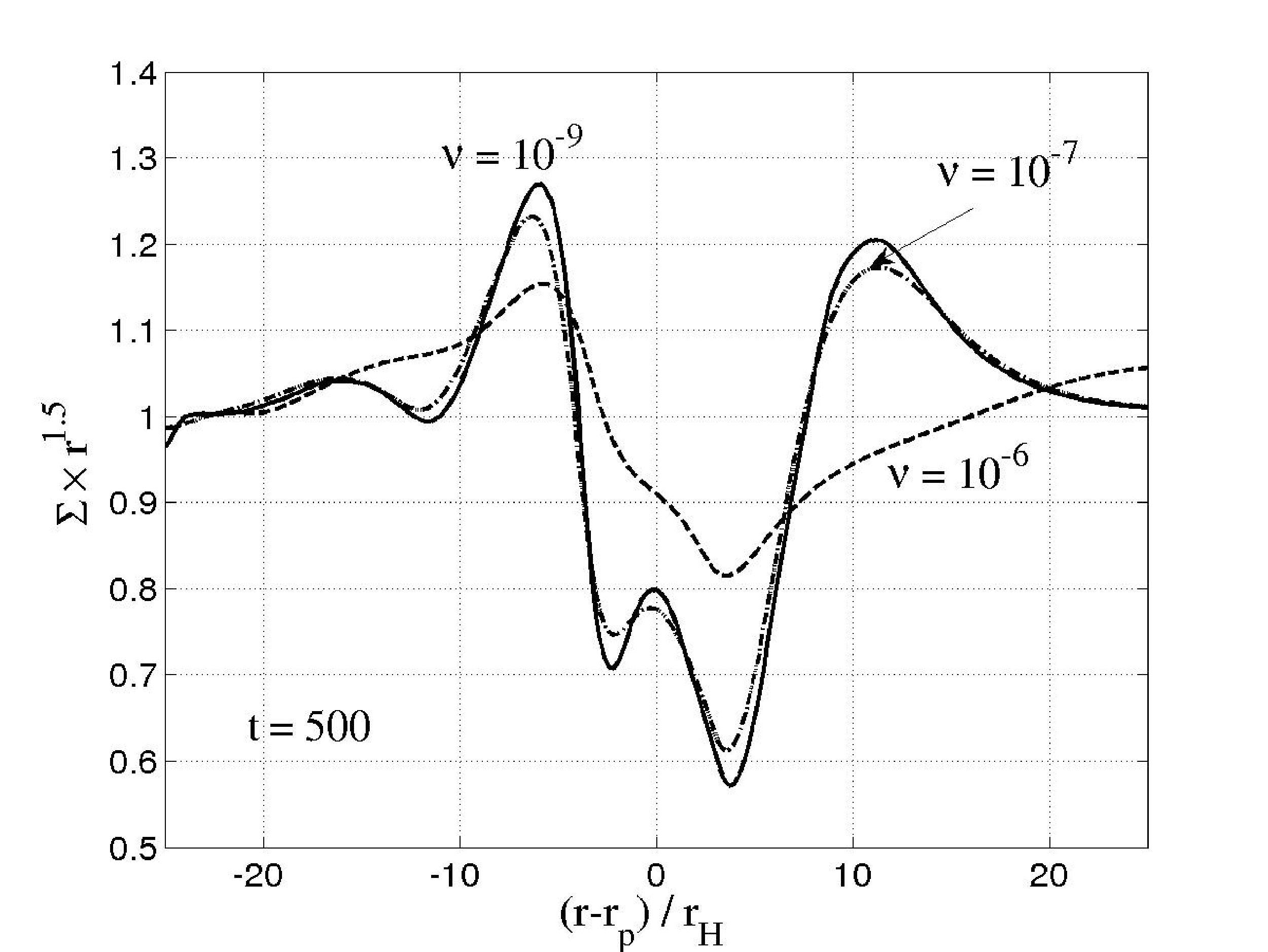}}
\caption{Azimuthally averaged disk surface density distribution ($\Sigma \times
  r^{1.5}$), normalized by its initial value at the orbital
  radius of the planet,
    at time $t=500 P$. 
    The horizontal axis is the radius relative the the planet radius 
  in units of the Hill radius.
  Model parameters are the same as in Fig. 1. The initial
  $\Sigma \times
  r^{1.5} $ is constant and equal to unity.
  For the lower viscosities, a gap is opening.
  \label{fig:rhos_vis_all}}
\end{figure}
\end{center}

Figure \ref{fig:vis_all} shows the influence of the imposed
disk viscosity on the
migration for a planet with $\mu = 3\times 10^{-5}$ or $10
M_{\oplus}$, $h = 0.035 $, and $f=5$.  For relatively large viscosity
($\nu = 10^{-6}$, $\alpha = 8 \times 10^{-4}$), the migration rates
agree well with the Type I rates given by Tanaka et al. (2002), as
discussed above. At early times the migration rates are largely
independent of the disk viscosity.  As viscosity decreases, after about 100
P, the migration is drastically slowed or completely halted. The rapid
oscillations with modest amplitude at $t \sim 800 P$ are due to
the excitation of vortices from a secondary instability (Koller et al.
2003; Li et al.  2005; see also Li et al.  2001), which will be a
subject for future studies. Figure \ref{fig:rhos_vis_all} reveals the
reason for the slow-down. A partial gap in the disk around the planet
has formed at $t=500 P$ and the density profile deviates significantly
from the initial power-law. The asymmetry in the density distribution
interior and exterior to the planet has reduced the contribution from
the outer Lindblad torque so that the net torque is approximately
zero.  We verified that the slow-down shown in Fig. \ref{fig:vis_all}
is largely caused by the density redistribution.  In principle, the
torque distributions per unit disk mass ($d T/dM(r)$ see Fig 1 in
D'Angelo \& Lubow 2008) could also be affected. 
But we find these changes only slightly
modify (at the few percent level) the net migration torque.

In the case of $\nu = 10^{-6}$ in Fig~\ref{fig:rhos_vis_all}, the
profile is qualitatively similar to the expectations of steady state
theory (Ward 1997; Rafikov 2002).  In particular there is a density
peak at $r < r_p$ and a trough at $r > r_p$.  The torque that the disk
exerts on the planet is localized to a region of a few times the disk
thickness or about $\pm 4 r_H$ from radius $r_p$.  In
Fig~\ref{fig:rhos_vis_all} for $\nu = 10^{-6}$ , we see that the
perturbed density extends over a somewhat greater region of space,
suggesting that nonlocal damping is involved.

In the case of lower viscosities, $\nu = 10^{-7}, 10^{-9} $ in
Fig~\ref{fig:rhos_vis_all}, the density profiles are quite different
from the $\nu = 10^{-6} $ case.  In these cases, the planet has
effectively stopped migrating and the planet is opening a gap. We find
that the gap deepens over time, i.e., the system does not reach a
steady state. The large density gradients cause the vortex instability
to develop, as discussed above.

\begin{center}
\begin{figure}
\centerline{\includegraphics[height=2in,width=3in,angle=0]{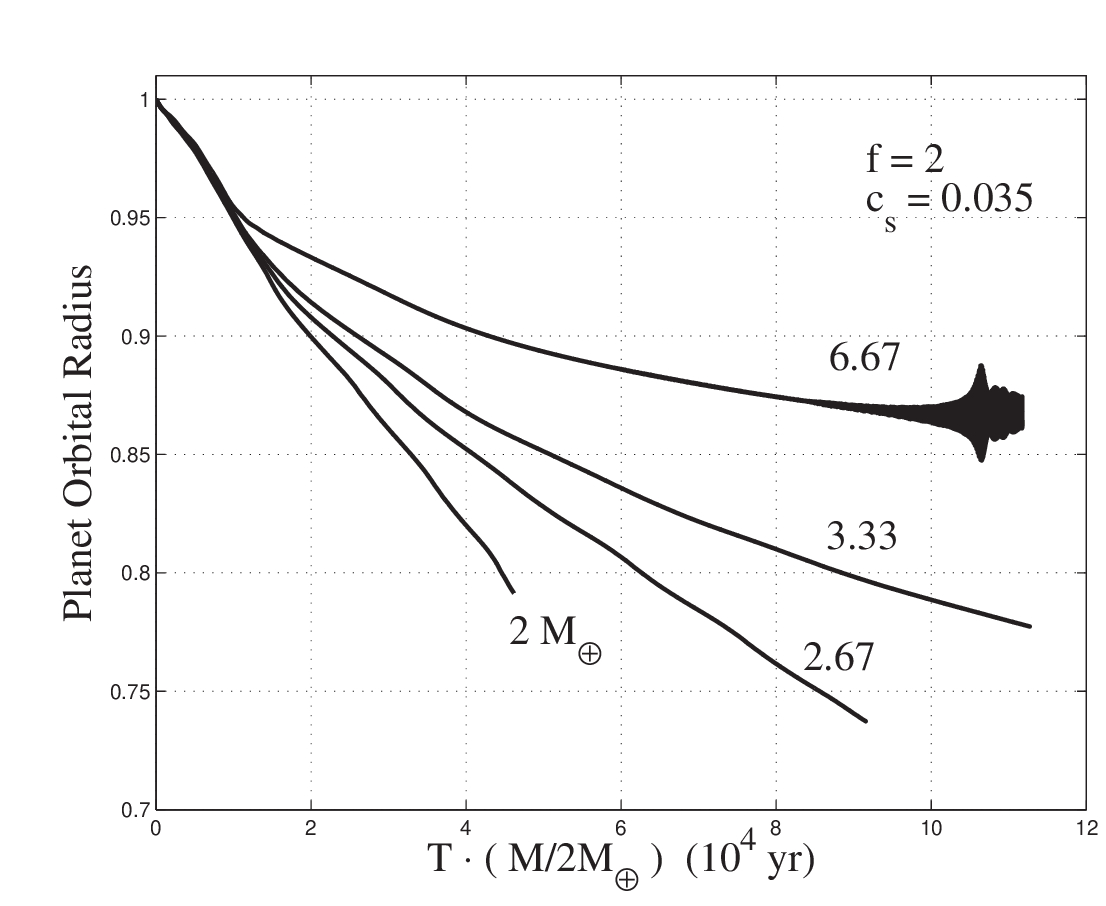}}
\centerline{\includegraphics[height=2in,width=3in,angle=0]{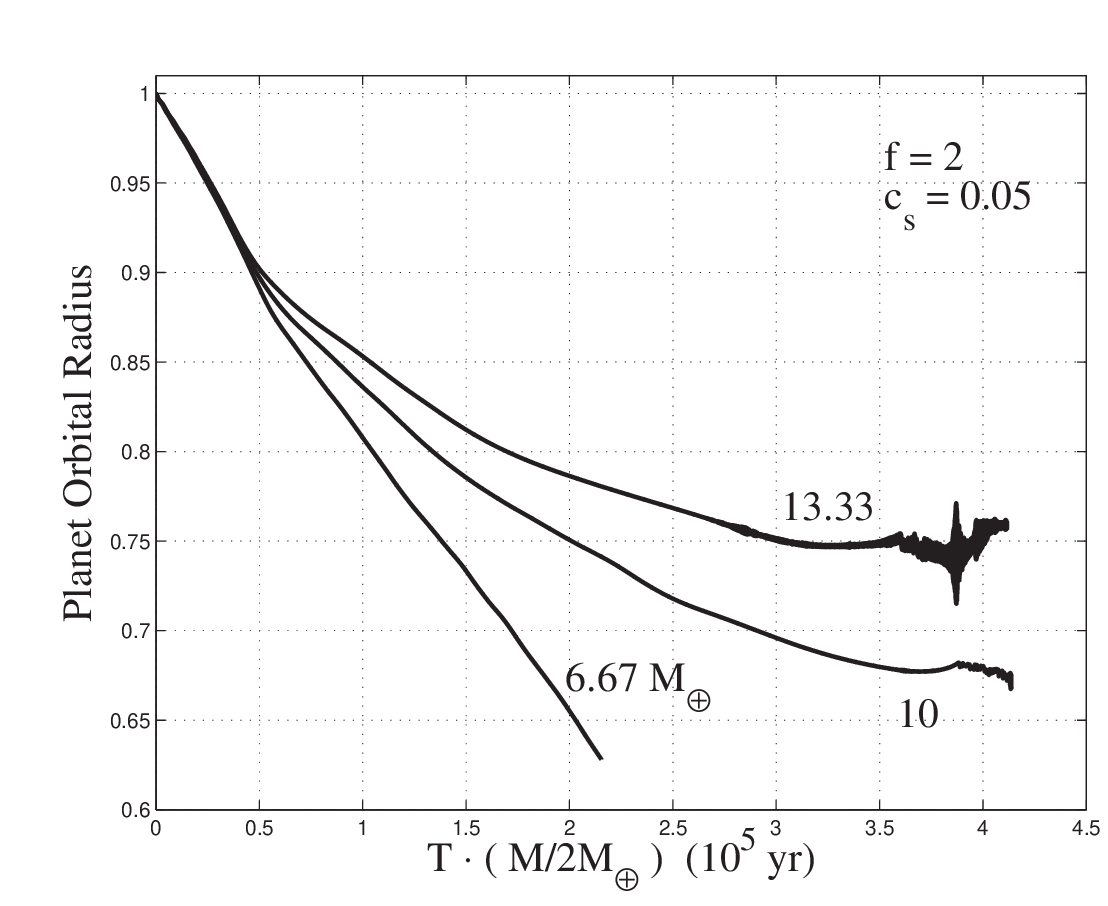}}
\caption{Migration history for several planet masses in nearly
  inviscid ($\nu = 0$) disks. 
  As the planet mass
  increases, its migration transitions from the Type I 
  to being much slower. 
  The vertical axis is the planet orbital radius in units of the
  initial  orbital
  radius.
  The horizontal axis is the time multiplied by the ratio of the planet
  mass  to $2 M_{\oplus}$. Each planet is assumed to have an initial orbital
  period of 11.87 yr.
  \label{fig:r_t_crit}}
\end{figure}
\end{center}

The strong density feedback in nearly inviscid disks and the reduction
in the total torque suggest the existence of critical planet mass
above which migration can be slowed significantly or halted.  Fig.
\ref{fig:r_t_crit} shows the transition from the usual Type I
migration to a much slower migration as the planet mass is increased,
for $f=2$, $h= 0.035, 0.05$, and $\nu = 0$.  The reduction in
migration rates is gradual, so it is difficult to define a single
value above which the migration will be halted. The values we
determine apply over the time range of several thousand $P$.  We have
performed a large number of runs for six different disk properties:
$f=1, 2, 5$, $h = 0.035$ and $0.05$. In Table 1 we give the estimates
of the planet masses in which the migration has significantly slowed
in our simulations. Above these values, migration was found to be
halted.

Previous studies have emphasized the role of gap formation in slowing
down the planet migration (Lin \& Papaloizou 1986; Crida \& Morbidelli
2007). From Fig. 2, we can see that a partial gap is formed at late
time. But the slowing down of the migration has started much earlier
(see also Fig. 3) where the gap is much less deep. This is consistent
with our interpretation that the slowing down of the migration is
primarily caused by the torque resulting from the asymmetric mass
re-distribution, i.e., the density feedback effects discussed at the
end of \S \ref{sec:intro}.  With the planet mass above the critical
mass, it is no longer possible to have steady state migration (as
explained in the analytic studies). In this regime, a gap gradually
develops and it deepens with time. But, it is still the asymmetry in
the density distribution (see Fig. 2) that ensures a much reduced (or
zero) net total torque.


\section{Discussions}
\label{sec:diss}

The local wave damping model of Ward \& Hourigan (1989) suggests
critical planet masses of $M_{\rm cr} \sim \Sigma r_p^2 h^3$, which
evaluates to $0.006 f M_{\oplus}$ and $0.02 f M_{\oplus}$ for $h =
0.035$ and $0.5$, respectively.  These values differ from Table 1 by a
factor of more than 100. However the scaling of the critical mass with
disk thickness is close to $h^3$ as given by this theory. The scaling
of $M_{\rm cr}$ with surface density in Table 1 is, however, weaker than
that suggested by this theory.  In another local wave damping model,
Ward (1997) suggests that $M_{\rm cr} \sim 0.2 \Sigma r_p^2 h$, which
evaluates to $1.1 f M_{\oplus}$ and $1.6 f M_{\oplus}$ for $h = 0.035$
and $0.05$, respectively. Although these values are numerically closer
to the simulation values in Table 1, the predicted linear scaling in
both $f$ and $h$ does not agree with the trends in the Table 1.
The analytic model of Rafikov (2002)
includes the effects of nonlocal damping by means of shocks.  In that
case, the critical masses are given by
\begin{equation}
M_{cr} = \frac{2 c_s^3}{3 \Omega G}~ {\rm min}\left[5.2 Q^{-5/7}, \,
3.8 ( Q/h )^{-5/13} \right],
\end{equation}
where $Q = \Omega c_s/(\pi G \Sigma)$.  
All the simulated cases correspond to the strong feedback
branch of $M_{cr}$ that is given by the second
argument of the $min$ function. 
Values for these critical
masses are also given in Table 1.  We see that the agreement between
the simulation and theory is quite good.

The critical masses are in the range of the core masses during the
slow phase of gas accretion in the core accretion model of planet
formation. This result suggests that planet migration might not be a
limiting factor in planet formation. The phase of run-away mass
accretion follows the slow evolution phase.  Previous studies suggest
that run-away mass accretion to $1 M_J$ in a disk with $\alpha \simeq
0.004$ occurs in about $10^{5} y$ (e.g., D'Angelo \& Lubow 2008).
During this phase, we speculate that
more modest levels of turbulent viscosity $ 10^{-4}
< \alpha < 10^{-3}$ may provide sufficient accretion to form a $1 M_J$
planet within $\sim 10^{6} y$, while remaining in this nearly laminar
disk regime of slow Type I migration.  

Although this picture is suggestive of a resolution of the migration
problem, the longer term evolution of nearly laminar disk-planet
systems requires further exploration.  The slowly migrating planet
will continue to create a deeper gap over time. The steepening density
gradients should lead to the vortex instability (Koller et al.  2003;
Li et al. 2005).  The consequences of the vortex instability should be
explored.

\acknowledgments 

The research at LANL is supported by a Laboratory Directed Research
and Development program. S.L. acknowledges support from NASA Origins
grant NNX07AI72G.


\clearpage

\begin{table*}
\begin{center}
\caption{Critical Planet Mass $M_{cr} (M_{\oplus})$ from
  Simulations and Theory (Rafikov 2002)}
\begin{tabular}{|c|c|c|c|c|c|c|}
\hline
  & \multicolumn{2}{c|} {$f = 1$} & \multicolumn{2}{c|} {$f = 2$}
& \multicolumn{2}{c|} {$f = 5$}\\
\hline
$h$ & simulation & theory & simulation & theory & simulation & theory \\
\hline
0.05 & $\sim 9$ & $ 8.4$ & $\sim 9$ & $ 10.9$  & $\sim 15$ & $ 15.5$\\
\hline
0.035 & $\sim 3$ & $2.9$ & $\sim 3$ & $3.7$ & $\sim 6$ & $5.3$ \\
\hline
\end{tabular}
\end{center}
\end{table*}


\end{document}